\documentstyle[aps,epsfig,prl,preprint]{revtex}

\def\ptsq   {P^{2}_{T}}

\def\kpmz   {K^{0}_{L} \rightarrow \pi^+\pi^-\pi^0}
\def\kppg   {K^{0}_{L,S} \rightarrow \pi^+\pi^-\gamma}
\def\ksppg  {K^{0}_{S} \rightarrow \pi^+\pi^-\gamma}
\def\klppg  {K^{0}_{L} \rightarrow \pi^+\pi^-\gamma}
\def\klppgs {K^{0}_{L} \rightarrow \pi^+\pi^-\gamma^*}
\def\gsee   {\gamma^* \rightarrow e^+e^-}
\def\kppee  {K^{0}_{L} \rightarrow \pi^+\pi^- e^+ e^-}

\def\kpip   {K^{0}_{L} \rightarrow \pi^+\pi^-}

\def\egcom  {E^{*}_{\gamma}}
\def\egcoms {E^{*2}_{\gamma}}

\begin{document}
\draft

\title{Study of the $\klppg$ Direct Emission Vertex}

\pagestyle{plain}

\author{
A.~Alavi-Harati$^{12}$,
T.~Alexopoulos$^{12}$,
M.~Arenton$^{11}$,
K.~Arisaka$^2$,
S.~Averitte$^{10}$,
A.R.~Barker$^5$,
L.~Bellantoni$^7$,
A.~Bellavance$^9$,
J.~Belz$^{10,\dagger}$,
R.~Ben-David$^7$,
D.R.~Bergman$^{10}$,
E.~Blucher$^4$, 
G.J.~Bock$^7$,
C.~Bown$^4$,
S.~Bright$^4$,
E.~Cheu$^1$,
S.~Childress$^7$,
R.~Coleman$^7$,
M.D.~Corcoran$^9$,
G.~Corti$^{11}$, 
B.~Cox$^{11}$,
M.B.~Crisler$^7$,
A.R.~Erwin$^{12}$,
R.~Ford$^7$,
A.~Glazov$^4$,
A.~Golossanov$^{11}$,
G.~Graham$^4$, 
J.~Graham$^4$,
K.~Hagan$^{11}$,
E.~Halkiadakis$^{10}$,
K.~Hanagaki$^8$,  
S.~Hidaka$^8$,
Y.B.~Hsiung$^7$,
V.~Jejer$^{11}$,
D.A.~Jensen$^7$,
R.~Kessler$^4$,
H.G.E.~Kobrak$^{3}$,
J.~LaDue$^5$,
A.~Lath$^{10}$,
A.~Ledovskoy$^{11}$,
P.L.~McBride$^7$,
P.~Mikelsons$^5$,
E.~Monnier$^{4,*}$,
T.~Nakaya$^7$,
K.S.~Nelson$^{11}$,
H.~Nguyen$^7$,
V.~O'Dell$^7$, 
M.~Pang$^7$, 
R.~Pordes$^7$,
V.~Prasad$^4$, 
C.~Qiao$^4$, 
B.~Quinn$^4$,
E.J.~Ramberg$^7$, 
R.E.~Ray$^7$,
A.~Roodman$^4$, 
M.~Sadamoto$^8$, 
S.~Schnetzer$^{10}$,
K.~Senyo$^8$, 
P.~Shanahan$^7$,
P.S.~Shawhan$^4$, 
W.~Slater$^2$,
N.~Solomey$^4$,
S.V.~Somalwar$^{10}$, 
R.L.~Stone$^{10}$, 
I.~Suzuki$^8$,
E.C.~Swallow$^{4,6}$,
S.A.~Taegar$^1$,
R.J.~Tesarek$^{10}$, 
G.B.~Thomson$^{10}$,
P.A.~Toale$^5$,
A.~Tripathi$^2$,
R.~Tschirhart$^7$, 
Y.W.~Wah$^4$,
J.~Wang$^1$,
H.B.~White$^7$, 
J.~Whitmore$^7$,
B.~Winstein$^4$, 
R.~Winston$^4$, 
T.~Yamanaka$^8$,
E.D.~Zimmerman$^4$
\begin{center} (KTeV Collaboration) \end{center}}
\address{
$^1$ University of Arizona, Tucson, Arizona 85721 \\
$^2$ University of California at Los Angeles, Los Angeles, California 90095 \\
$^{3}$ University of California at San Diego, La Jolla, California 92093 \\
$^4$ The Enrico Fermi Institute, The University of Chicago,
Chicago, Illinois 60637 \\
$^5$ University of Colorado, Boulder, Colorado 80309 \\
$^6$ Elmhurst College, Elmhurst, Illinois 60126\\
$^7$ Fermi National Accelerator Laboratory, Batavia, Illinois 60510 \\
$^8$ Osaka University, Toyonaka, Osaka 560 Japan \\
$^9$ Rice University, Houston, Texas 77005 \\
$^{10}$ Rutgers University, Piscataway, New Jersey 08855 \\
$^{11}$ The University of Virginia, Charlottesville, Virginia 22901 \\
$^{12}$ University of Wisconsin, Madison, Wisconsin 53706 \\
}

\maketitle


\vspace{-12.0cm}
\hspace{14cm}
\large (Rutgers--00--17) \normalsize
\vspace{12.0cm}
\hspace{-14cm}

\vspace{-0.3cm}
\begin{abstract}
\noindent
We have perfomed studies of the $\klppg$ Direct Emission 
(DE) and Inner Bremsstrahlung (IB) vertices, based on data collected 
by KTeV during the 1996 Fermilab fixed target run. We find
$a_1/a_2 = -0.737 \pm 0.034$ GeV$^2$ for the DE form--factor parameter
in the $\rho$--propagator parametrization, and report on fits of
the form--factor to linear and quadratic functions as well. 
We concurrently measure  
$\Gamma(\klppg,\egcom >20 \hspace{0.2cm} {\rm MeV})/\Gamma(\kpip) = 
 (20.8 \pm 0.3) \times 10^{-3}$, and a $\klppg$ DE/(DE + IB)
branching ratio of $0.683 \pm 0.011$.
\end{abstract}
\pacs{PACS numbers: 11.30.Er, 13.25.Es, 13.40.Hq, 14.40.Aq}

%

\begin{figure}[ht]
\centerline{ \epsfxsize 1.0 truein \epsfbox{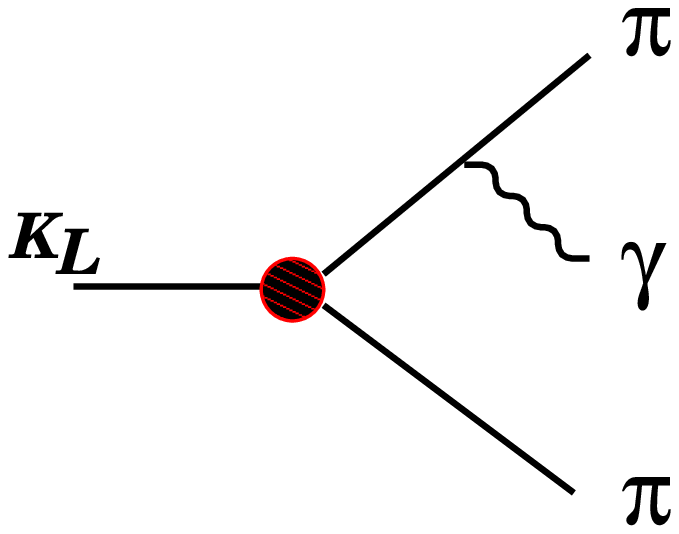} 
             \hspace{1.0cm}  
             \epsfxsize 1.0 truein \epsfbox{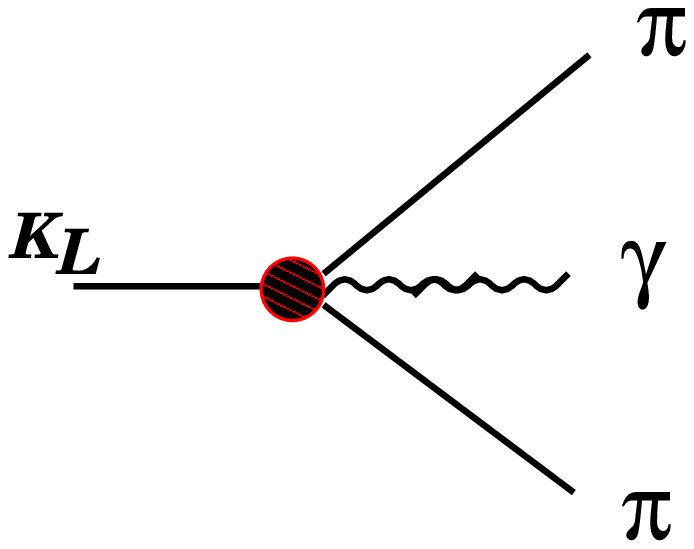} }
\vspace{0.5cm}
\caption{ \label{diagrams} The decay $\klppg$ arises primarily from
	  the contributions of the Electric Dipole (E1) ``Inner 
          Bremsstrahlung'' (left) and Magnetic Dipole (M1) 
          ``Direct Emission'' (right) diagrams above.} 
\end{figure}

The decay $\klppg$ (Figure~\ref{diagrams}) is a potential
new window into the phenomenon of CP--violation~\cite{cos67,seh67,dam96}. 
This decay arises primarily from the
CP--violating Electric Dipole (E1) ``Inner Bremsstrahlung'' (IB) and
the CP--conserving  Magnetic Dipole (M1) ``Direct Emission'' (DE)
processes. KTeV recently reported~\cite{ala00} the observation of a 
CP--violating angular asymmetry between the $\pi^+\pi^-$ and $e^+e^-$ 
decay planes in the $\klppgs$, $\gsee$ mode, arising from the interference
of DE and IB photon polarization states~\cite{seh92,hei93,elw95}. 
If an E1 DE term were also present, this would generate a direct 
CP--violating effect.

Understanding the precise photon energy ($E^*_{\gamma}$) spectrum
from the DE amplitude is crucial to interpretation of CP--violating 
effects and may shed light on particular chiral 
models~\cite{lin88}. Previous experiments~\cite{car80,ram93} have 
observed evidence for an energy shift in the DE $E^*_{\gamma}$ spectrum, 
interpreted as evidence for an $E^*_{\gamma}$--dependent form--factor 
modification to the pure--M1 DE amplitude. 
Here we report on the first measurement of this form--factor from direct
fits to the data, using the rare decay mode $\klppg$.

We consider two separate form--factor parametrizations.
Historically, the $\rho$--propagator form~\cite{lin88}
\begin{equation}
{\cal F} = \frac{a_1}{(m^2_{\rho}-m^2_K)+2m_{K}E^*_{\gamma}} + a_2
\end{equation}
has been used, where $a_1/a_2$ is the quantity of interest.
More generally the form factor may be expressed as a Taylor
series in $\egcom$
\begin{equation}
{\cal F} = \left(1 + \frac{r\egcom}{M_K} + \frac{s\egcoms}{M^2_K} 
        + \cdots \right)
\end{equation}
where the interesting quantitites are the slope 
$r$ and the quadratic parameter $s$. Recent theoretical 
work~\cite{eck94,val93} has attempted to calculate
these parameters. 


The data presented here was collected by KTeV operating in the
E832 configuration~\cite{ala99} during the 1996 Fermilab fixed 
target run. A proton beam of intensity $\sim 3 \times 10^{12}$ protons 
per 19 second spill incident at an angle of 4.8 mr on a BeO target
was employed to produce two nearly parallel $K^0_L$ beams. In
E832, one of these beams was incident on an active $K^0_S$ 
regenerator. Data collected in the regenerator beam was ignored 
for this analysis, except as a check of our understanding of the 
IB decay amplitude. The configuration of the KTeV E832 spectrometer
consisted of a vacuum decay region, a magnetic spectrometer
with four drift chambers, photon vetoes, a trigger scintillator
bank, a CsI electromagnetic calorimeter, and a muon veto bank.  

Signal $\klppg$ and normalization $\kpip$ candidates were selected from 
the two--charged--track trigger. Offline, the sample was further purified 
by requiring the presence of two reconstructed tracks with a good 
vertex--$\chi^2$ within the fiducial aperture of the detector, and no 
significant activity in the photon veto counters. The tracks' kinematics were 
required to be inconsistent with $\Lambda^0 \rightarrow p^+ \pi^-$ decays. 
The energy deposited by the charged pions in the calorimeter
was required to be less than $0.85 \times$ the spectrometer momentum in
order to eliminate backgrounds from $K_{e3}$ decays.  
Candidate events were required to have a decay vertex 
between 110 and 156 meters downstream of the target and a total energy 
between 20 and 160 GeV. 

$\klppg$ candidates were subject to the additional criterion of 
requiring that the quantity
\begin{equation}
P^2_{\pi^0} \equiv
\frac{(M^2_K - M^2_{\pi^0} - M^2_{\pi\pi})^2 - 4M^2_{\pi^0}M^2_{\pi\pi} 
  - 4M^2_K(P^2_T)_{\pi\pi}}
{4((P^2_T)_{\pi\pi} + M^2_{\pi\pi})}
\end{equation}
be negative, {\em i.e.} by requiring that the $\pi^0$ momentum
be imaginary under a $\kpmz$ hypothesis. This cut virtually eliminates 
the $\kpmz$ background to the $\klppg$ event sample. In addition, at least 
one non--track--associated cluster in the calorimeter was required to 
possess $\geq 1.5$ GeV of energy, and to be at least 3 cm removed from 
the calorimeter edges. This ``photon cluster'' had to be at least 20 cm
from the nearest track projection to reject background from $\kpip$
events accompanied by pion hadronic showers in the calorimeter. The 
photon was required to have an energy of at least 20 MeV in the 
three--body center of momentum.

\begin{figure}[ht]
\centerline{ \epsfxsize 1.75 truein \epsfbox{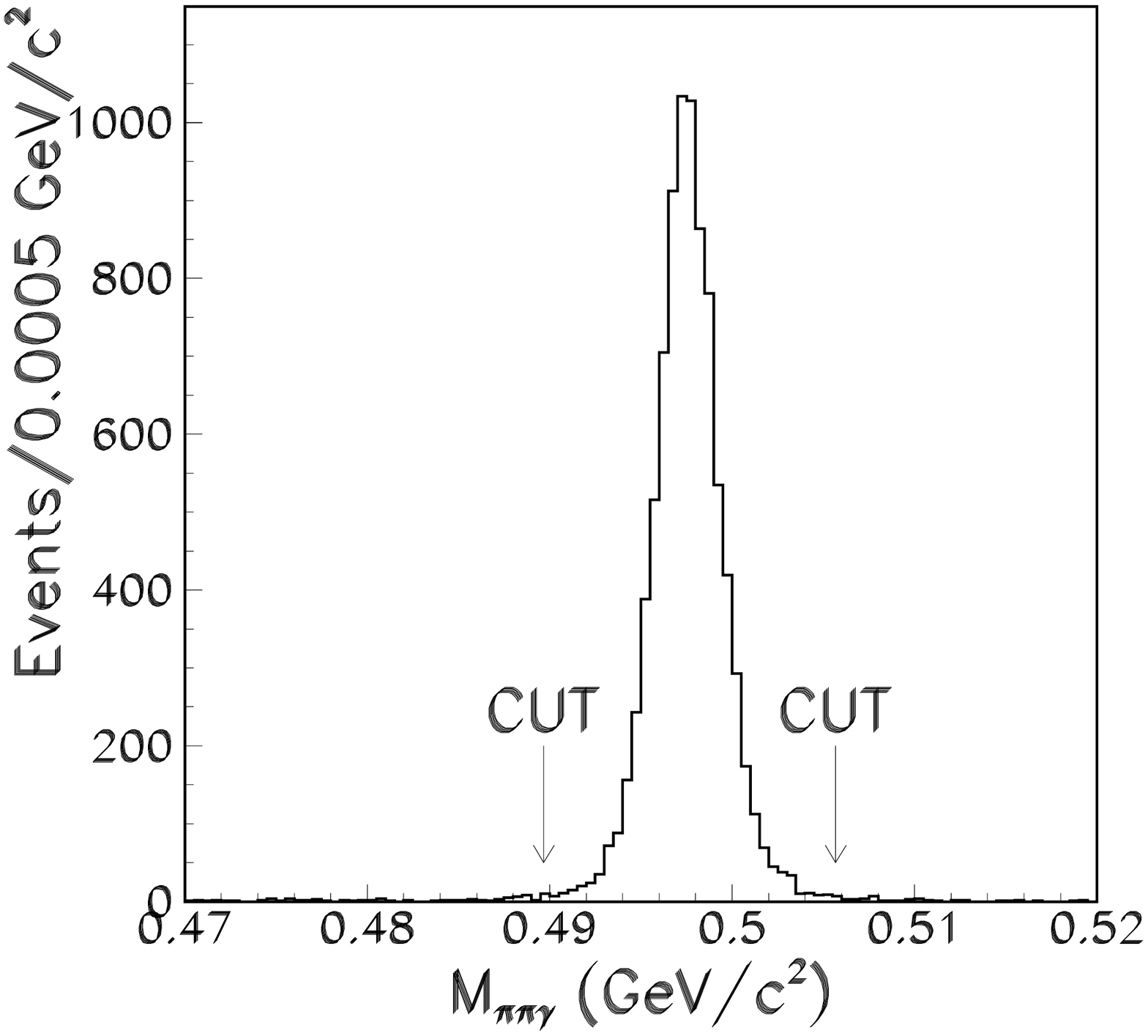} 
             \epsfxsize 1.75 truein \epsfbox{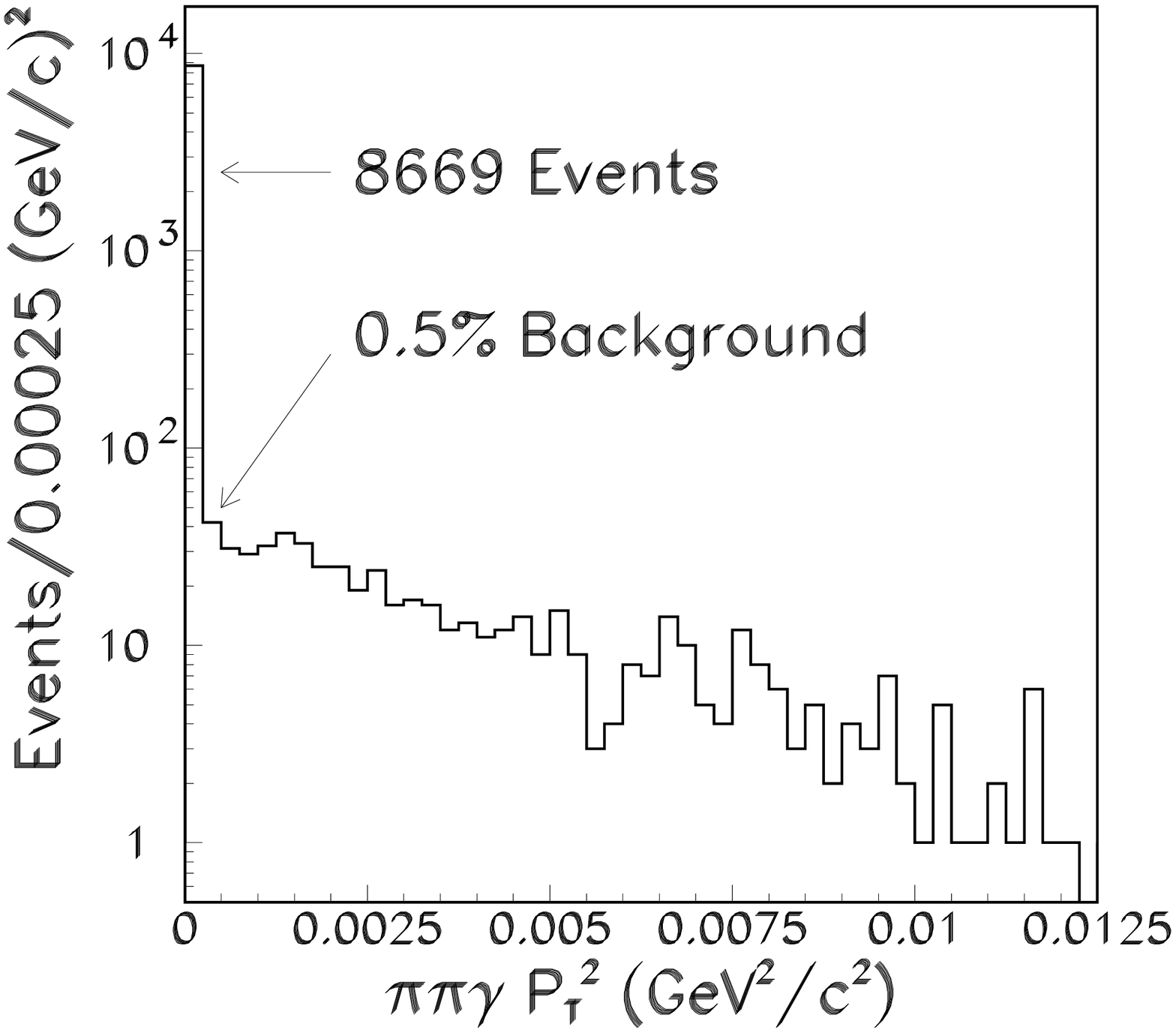} }
\vspace{0.5cm}
\caption{ \label{signal} 
a) $M_{\pi\pi\gamma}$ distribution of candidate $\klppg$ events, 
   all other cuts applied. Arrows indicate final cuts
   at $M_k \pm 8$ MeV/c$^2$.
b) Kaon transverse momentum squared ($\ptsq$) distribution of 
   candidate $\klppg$ events, all other cuts applied. The cut on 
   this quantity requires events to be within the first bin
   (0.000250 GeV$^2$/c$^2$) on this plot.
} 
\end{figure}

Figures~\ref{signal} (a) and (b) show the vacuum beam $\kppg$ data 
after final cuts. A very clean signal of 8,669 $\kppg$ events is
achieved, with a background of about 0.5\%. 
Approximately 0.6\% of the 
events in the peak in Figure~\ref{signal} are residual $\ksppg$
decays from $K^0_S$'s generated at the target. A total of 4,482,706 
$\kpip$ events were accumulated for the normalization sample with 
a 0.1\% background.

\begin{figure}[ht]
\begin{center}
   {\epsfxsize=3.5in\epsffile{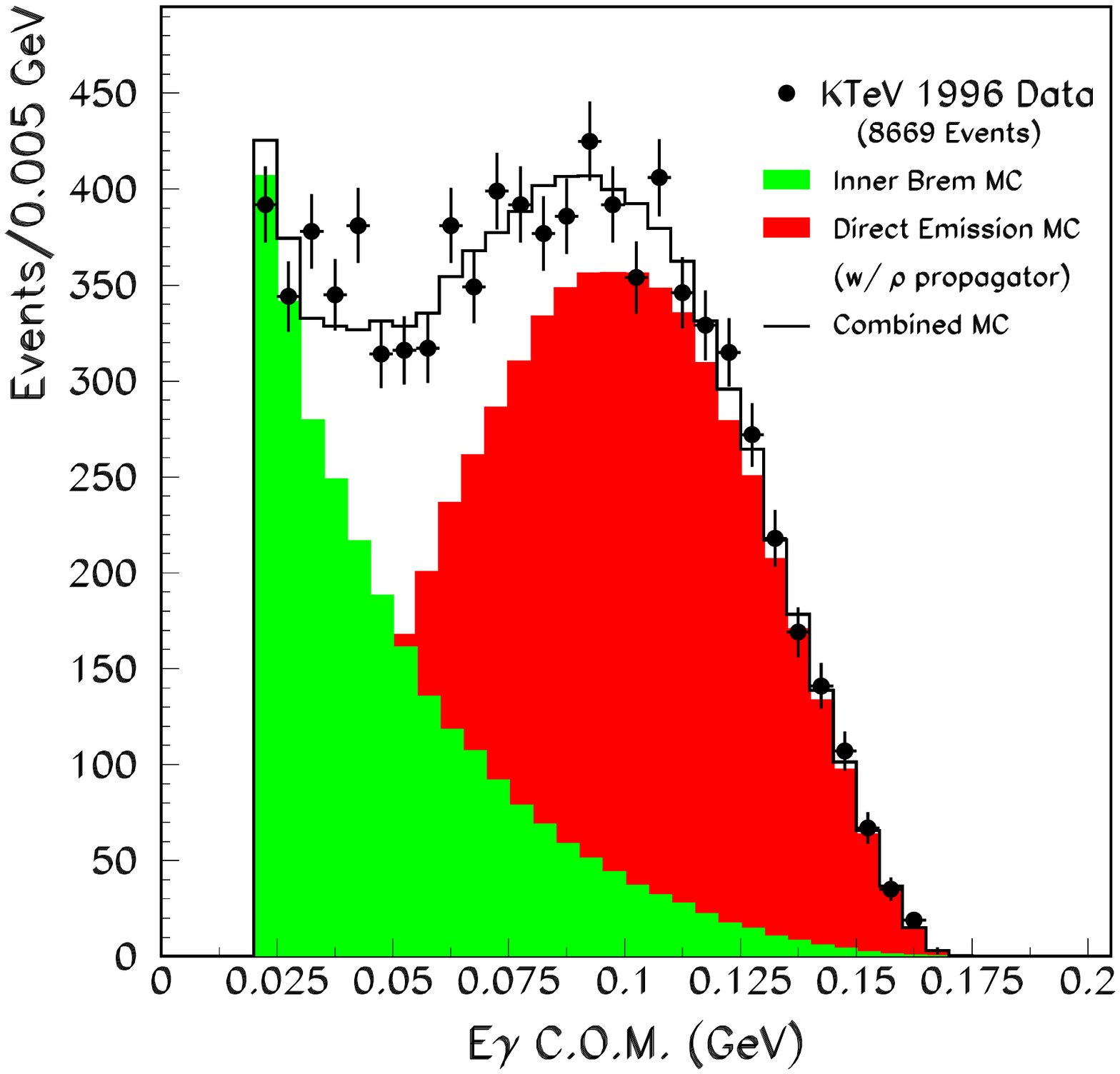}}
\end{center}
\caption{\label{egcom} $\klppg$ Monte Carlo/data overlay of photon energy
 distribution in the center of mass, for the best fit result. Shown also 
 (shaded) are the expected distributions for pure
 E1 Inner--Bremsstrahlung and form--factor--modified M1 Direct Emission.
 The ``combined'' Monte Carlo plot shown assumes these two are the only 
 contributions to the decay.}
\end{figure}

To extract the Direct Emission form--factor, we consider the distribution 
in $\egcom$, the photon energy in the center of momentum (Figure~\ref{egcom}). 
We wish to extract the relative contributions of the Inner Bremsstrahlung 
and DE terms, as well as the energy shift of the DE spectrum due to the 
presence of a form factor. We assume that E1 IB and form--factor--modified 
M1 DE are the only significant contributions to the decay. The possibility 
of an interference term arising from the presence of an E1 DE contribution
is also briefly discussed below. 

We perform a MINUIT~\cite{minuit} $\chi^2$--minimization 
fit to combine Monte Carlo DE and IB $\egcom$ distributions, 
and extract the relative DE and IB contributions to the data 
as well as the DE form--factor. In the fit, the form factor 
parameters ($a_1/a_2$ or $r$ and $s$) and the ratio 
\begin{equation}
f = \frac{\Gamma_{DE}}{(\Gamma_{DE}+\Gamma_{IB})}
\end{equation}
are allowed to float simultaneously. A $\chi^2$ is formed from comparison
of the resultant $DE + IB$ summed Monte Carlo histogram with the data, 
and minimized to obtain the best fit result. 

\begin{figure}[ht]
\begin{center}
   {\epsfxsize=3.5in\epsffile{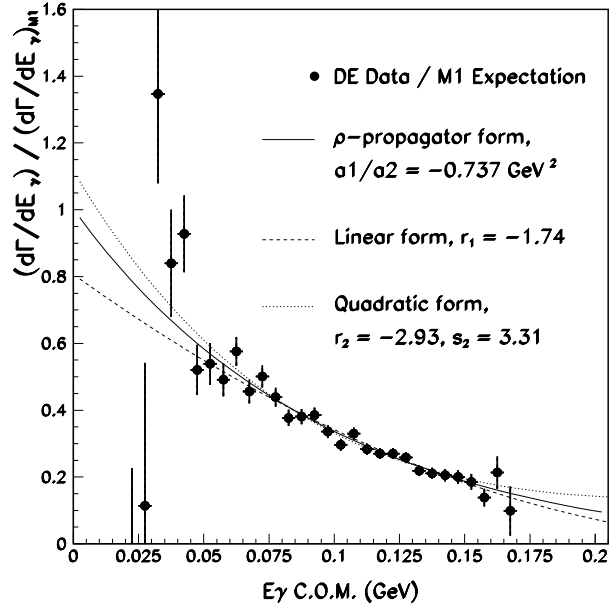}}
\end{center}
\caption{\label{ffcomp} 
Ratio (points) of (IB--Subtracted) Direct--Emission data to the 
expectation for a pure M1 $\egcom$ spectrum. Vertical scale is 
arbitrary. Shown for comparison are the $\rho$--propagator (solid), 
linear (dashed), and quadratic (dotted) form--factor parametrizations. 
A modification to the pure--M1 spectrum is clearly supported by the data.}
\end{figure}

Table~\ref{ppgphy} summarizes
the numerical fitting results for the three form-factor parametrizations. 
For the Taylor series parametrization, fits were performed with the 
quadratic parameter both fixed at zero (floating a single parameter, 
$r_1$) and allowed to vary (floating two parameters, $r_2$ and $s_2$). 
Data--Monte Carlo agreement in the $E^*_{\gamma}$ distribution is shown 
in Figure~\ref{egcom}, for the $\rho$--propagator form. 
Figure~\ref{ffcomp} illustrates the expected effect of the various 
parametrizations on the pure M1 direct emission spectrum. 

We see in the data presented in Table~\ref{ppgphy} and Figure~\ref{ffcomp}
clear evidence for a modification to the pure--M1 DE spectrum.  
All fits are good, though the $\rho$--propagator hypothesis stands 
out slightly: It gives the best $\chi^2$ for a single--parameter fit, 
and the two--parameter fit results are in good agreement with the values 
$r_2 = -2.70$ and $s_2 = 3.87$ obtained by Taylor--expanding the 
$\rho$--propagator form. The size of this energy shift, in particular the 
need to take into account terms of second--order in $\egcom$, is not 
currently understood within the chiral perturbation theory model for 
this decay~\cite{eck94}.

Additional contributions to the photon energy spectrum are expected
from CP--violating higher--order multipole contributions to the 
Direct Emission amplitude. One possible consequence of these multipole
terms is the presence of a charge--asymmetry in the $\pi^+$ versus $\pi^-$ 
Dalitz plot~\cite{cos67}. 
We exclude asymmetries larger than 2.4\% at 90\% c.l. with the present 
data. 

One might also expect a contribution to the $E^*_{\gamma}$
spectrum from CP--violating E1 DE, which could interfere 
with the E1 IB. We have allowed for such a term, by performing
a separate fit in which we assumed an E1 DE amplitude constant in 
$E^*_{\gamma}$, and searched for the corresponding interference 
contribution~\cite{val93} to the $\egcom$ spectrum. The form--factor
parameter and $f$ are allowed to float simultaneously, while fixing
the $IB$ rate at its theoretical value of $7.00 \times 10^{-3}$~\cite{cos67}.
Based on this fit, we set an upper limit of 
$\Gamma_{IN}/(\Gamma_{DE}+\Gamma_{IB}+\Gamma_{IN})
 \leq 0.30 \hspace{0.2cm} {\rm (90\% \hspace{0.1cm} c.l.)}$
on the contribution to the decay rate of the CP--violating 
interference term. 
$\Gamma_{IN}/\Gamma_{ALL} = 0.30$ corresponds to a 22\% decrease in
the $\Gamma_{DE}/\Gamma_{\rm ALL}$ ratio. 

Background subtraction under the $E^*_{\gamma}$ distribution (with 
shape determined by study of $P^2_T$ sideband) resulted in no 
statistically significant change in either the form--factor parameter 
or the $DE/(DE+IB)$ ratio. Accidental activity in the detector 
was also found to have no statistically significant effect.  
The form factor parameters are found to have a slight (1.9\%) 
sensitivity to variations of the lower $\egcom$ cutoff 
(nominally 20 MeV) used in the fit. We also assign systematic 
uncertainties of 1.8\% due to detector acceptance, and 1.4\% 
due to the effects of uncertainties in the calorimeter photon 
energy scale. Bias due to Level 3 trigger inefficiency
in the 1996 charged data set~\cite{ala99,pss99} is not significant 
in this analysis at the present level of statistics.  

Note (Table~\ref{ppgphy}) that the measured $DE/(DE+IB)$ ratio is 
insensitive to the particular choice of form--factor parametrization. 
Uncertainty in the detector acceptance contributes a 0.7\% systematic
uncertainty to this ratio, while variations in analysis cuts contribute
$0.6$\%. The systematic uncertainty due to the calorimeter photon
energy scale is $0.2$\%.

We find the absolute $\klppg$ branching ratio by normalizing the 
signal to the $\kpip$ channel. We determine the ratio of $\kpip/\klppg$ 
acceptances using a Monte Carlo simulation of the full detector and 
offline analysis criteria. Based on this, we calculate  
\begin{equation}
 \frac{\Gamma(\klppg)}{\Gamma(\kpip)} = (20.8 \pm 0.2 \pm 0.2 ) \times 10^{-3}
\label{eq:finalbr}
\end{equation}
(where the first error is statistical and the second systematic) 
for $\klppg$ events with $\egcom > 20 \hspace{0.1cm} {\rm MeV}$. The
systematic uncertainty is due primarily to the effects of $K^0_S$ 
contamination in the vacuum beam.
Assuming no contribution from interference and 
using the $DE/(DE+IB)$ result from Table~\ref{ppgphy}, 
we obtain the final $\pi^+\pi^-$--normalized branching ratios
$(14.2 \pm 0.2 \pm 0.2 ) \times 10^{-3}$ for Direct Emission and 
$(6.6 \pm 0.2 \pm 0.2 ) \times 10^{-3}$ for Inner Bremsstrahlung.


Our measured IB branching ratio is consistent with both the
Q.E.D. prediction of $7.00 \times 10^{-3}$ and
the most recent experimental result of FNAL E731~\cite{ram93};
$(7.3 \pm 0.4) \times 10^{-3}$, based on a sample of 3,136 $\klppg$
events. The $DE/(DE+IB)$ ratio is also in good agreement with the
E731 result of $0.685 \pm 0.041$, although the DE branching ratio
differs from the E731 result of $(15.7 \pm 0.7) \times 10^{-3}$
by 1.8 standard deviations. The present form--factor results differ
significantly from that reported by E731 ($a_1/a_2 = -1.8 \pm 0.2$
GeV$^2$) but this discrepancy has been understood: The E731 $a_1/a_2$
form--factor was {\em inferred}~\cite{ram99} from the model of Lin and
Valencia~\cite{lin88}, on the basis of the measured $DE$ branching ratio,
whereas our results are obtained {\em directly} by performing fits to the
$\egcom$ distribution. The underlying datasets in the two experiments are
consistent with each other, and a reanalysis of the E731 data using our  
method yields results consistent with ours. The KTeV $\klppg$ DE
form--factor result is also in good agreement with the result
$a_1/a_2 = -0.720 \pm 0.029$ GeV$^2$ extracted from the independent
$\kppee$ analysis from the KTeV E799 data set~\cite{ala00}.

In conclusion, we have made the first direct measurements 
of the $\klppg$ Direct Emission form--factor, including  
$a_1/a_2 = -0.737 \pm 0.034$ GeV$^2$. We find no evidence 
for new CP--violating effects in the photon energy spectrum.
Finally, we have made improved measurements of the 
DE ($(14.2 \pm 0.2 \pm 0.2 ) \times 10^{-3}$) and 
IB ($ (6.6 \pm 0.2 \pm 0.2 ) \times 10^{-3}$)
$\klppg$ branching ratios, normalized to the $\kpip$ channel. 

We thank German Valencia and Jusak Tandean for discussions concerning 
this work. We gratefully acknowledge the support of the technical staff 
of Fermilab and participating institutions. This work was supported 
in part by the U.S. DOE, NSF, and The Ministry of Education and 
Science of Japan. 



\begin{table}[ht]
\begin{center}
\begin{tabular}{|c|c|c|c|} 
 Quantity   & $\rho$--propagator                     & Linear                             & Quadratic                       \\
\hline 
$\chi^2$/DOF& 38.8/27                                & 43.2/27                            & 37.6/26                         \\
 $a_1/a_2$  & $(-0.737 \pm 0.026 \pm 0.022)$ GeV$^2$ & ---                                & ---                             \\
 $r$        &  ---                                   & $-1.739 \pm 0.062 \pm 0.052$       & $-2.93 \pm 0.41 \pm 0.34$       \\
 $s$        &  ---                                   & ---                                & $ 3.31 \pm 1.15 \pm 0.96$       \\
 $f$        &  $0.683 \pm 0.009 \pm 0.007$           &  $0.682 \pm 0.009 \pm 0.007$       &  $0.684 \pm 0.011 \pm 0.007$    \\
\end{tabular}
\caption{\label{ppgphy} Summary of $\klppg$ physics results
  for the three form--factor parametrizations. First uncertainty
  is statistical, second systematic.} 
\end{center}
\end{table}

\end{document}